\documentstyle[multicol,aps]{revtex}
\renewcommand{\narrowtext}{\begin{multicols}{2}
\global\columnwidth20.5pc\noindent}
\renewcommand{\widetext}{\end{multicols}
\global\columnwidth42.5pc}
\begin{document}
\draft
\preprint{25 October 1997}
\title{Thermodynamic Properties of Heisenberg Ferrimagnetic
       Spin Chains:\\
       Ferromagnetic-Antiferromagnetic Crossover}
\author{Shoji Yamamoto}
\address{Department of Physics, Faculty of Science, Okayama University,\\
         Tsushima, Okayama 700, Japan}
\author{Takahiro Fukui}
\address{Institute of Advanced Energy, Kyoto University,
         Uji, Kyoto 611, Japan}
\date{Received \hspace{6cm}}
\maketitle
%-------------------------------------------------------------------------
%                        ABSTRACT
%-------------------------------------------------------------------------
\begin{abstract}
   We study thermodynamic properties of the one-dimensional Heisenberg
ferrimagnet with antiferromagnetically exchange-coupled two kinds of
spins $1$ and $1/2$.
The specific heat and the magnetic susceptibility are calculated
employing a modified spin-wave theory as well as a quantum Monte Carlo
method.
The specific heat is in proportion to $T^{1/2}$ at low enough
temperatures but shows a Schottky-like peak at mid temperatures.
The susceptibility diverges as $T^{-2}$.
We reveal that at low temperatures the model is regarded as a
ferromagnet, while at mid temperatures it behaves like a gapped
antiferromagnet.
\end{abstract}

\narrowtext

%-------------------------------------------------------------------------
%                        INTRODUCTION
%-------------------------------------------------------------------------
   A great progress has been made in studying the qualitative
difference \cite{Hald1} between the integer-spin and the
half-odd-integer-spin Heisenberg antiferromagnets. Recently there 
has appeared brand-new attempts to explore the quantum behavior of
mixed-spin chains with two kinds of spins.
Low-energy properties of various mixed-spin chains with singlet ground
states were analyzed\cite{Fuku1} via the nonlinear-$\sigma$-model
technique with particular emphasis on the competition between the
massive and the massless phases.
Mixed-spin chains with magnetic ground states serve us with another
topic and have in fact attracted much current interests
\cite{Alca1,Pati1,Breh1,Yama1,Nigg1,Yama2}.
Since we expect a gapless excitation from the ferrimagnetic ground
state, we there take little interest in the most naive problem
whether the spectrum is gapped or gapless.
Performing a numerical investigation by the use of conformal
invariance, Alcaraz and Malvezzi \cite{Alca1} indeed predicted the
appearance of quadratic dispersion relations for the mixed-spin
Heisenberg ferrimagnets, which is consistent with a spin-wave
calculation \cite{Pati1,Breh1}.
Actually the quadratic dispersion has explicitly been visualized
employing a quantum Monte Carlo (QMC) technique \cite{Yama1} and an
exact-diagonalization method \cite{Yama2}.
Thus we may expect quantum ferrimagnets to behave like ferromagnets
at low temperatures.
However, several authors \cite{Pati1,Breh1} have recently
reported that quantum ferrimagnets have a nontrivial excitation branch
gapped from the ground states as well as a gapless one, which
stimulates us to investigate their thermodynamic properties.
We show in this article that {\it the two distinct low-lying
excitations result in a novel temperature dependence of the thermal
quantities displaying both ferromagnetic and antiferromagnetic aspects}.
The present study sounds more fascinating considering
that all the mixed-spin-chain compounds synthesized so far exhibit
ferrimagnetic ground states \cite{Kahn1}.

%-------------------------------------------------------------------------
%                          BASIC PROPERTIES
%-------------------------------------------------------------------------
   We consider alternatively aligned two kinds of spins $S$ and $s$ on
a ring with antiferromagnetic exchange coupling between nearest
neighbors, which are described by the Hamiltonian
%-------------------------------------------------------------------------
%   hamiltonian
%-------------------------------------------------------------------------
\begin{equation}
   {\cal H}
      =J\sum_{j=1}^N
        \left(
         \mbox{\boldmath$S$}_{j} \cdot \mbox{\boldmath$s$}_{j}
        +\mbox{\boldmath$s$}_{j} \cdot \mbox{\boldmath$S$}_{j+1}
        \right)
      -g\mu_{\rm B}HM\,,
   \label{E:H}
\end{equation}
where
$M\equiv S^z+s^z\equiv\sum_{j=1}^N(S_j^z+s_j^z)$ is the total
magnetization, $N$ the number of the unit cells, $\mu_B$ the Bohr
magneton, and we have set the $g$ factors of the spins $S$ and $s$
equal to $g$.
For the sake of argument, we assume throughout the manuscript that
$S>s$.
Applying the Lieb-Mattis theorem \cite{Lieb1} to the Hamiltonian
(\ref{E:H}) with no external field, we find $(S-s)N$-fold degenerate
ground states.
Therefore the model exhibits ferrimagnetism instead of
antiferromagnetism.
The gapless and the gapped excitations, respectively, lie in the
subspaces of $M<(S-s)N$ and $M>(S-s)N$ and thus may be regarded as
ferromagnetic and antiferromagnetic.
In the case of $(S,s)=(1,1/2)$, the gap to the antiferromagnetic branch
was numerically estimated to be $1.759J$ \cite{Breh1,Yama2}.

%-------------------------------------------------------------------------
%                     NUMERICAL CALCULATION
%-------------------------------------------------------------------------
   Now we present QMC calculation of the thermal quantities at zero
field.
The recent field-theoretical argument \cite{Alca1} and density-matrix
renormalization-group (DMRG) study \cite{Pati1} both suggest that the
low-temperature properties of the model are qualitatively the same
regardless of the values of $S$ and $s$ as long as they differ from
each other.
That is why we restrict our numerical investigation to the case of
$(S,s)=(1,1/2)$.
We employ the QMC method based on the Suzuki-Trotter decomposition
\cite{Suzu1} of checkerboard type \cite{Hirs1} and its numerical
procedure has been detailed elsewhere \cite{Yama3}.
We mainly calculate the $N=32$ chain, which is long enough to discuss
the bulk properties.
Since the correlation length of the system is smaller than the length
of the unit cell \cite{Pati1,Breh1}, the thermal quantities show no
significant size dependence.

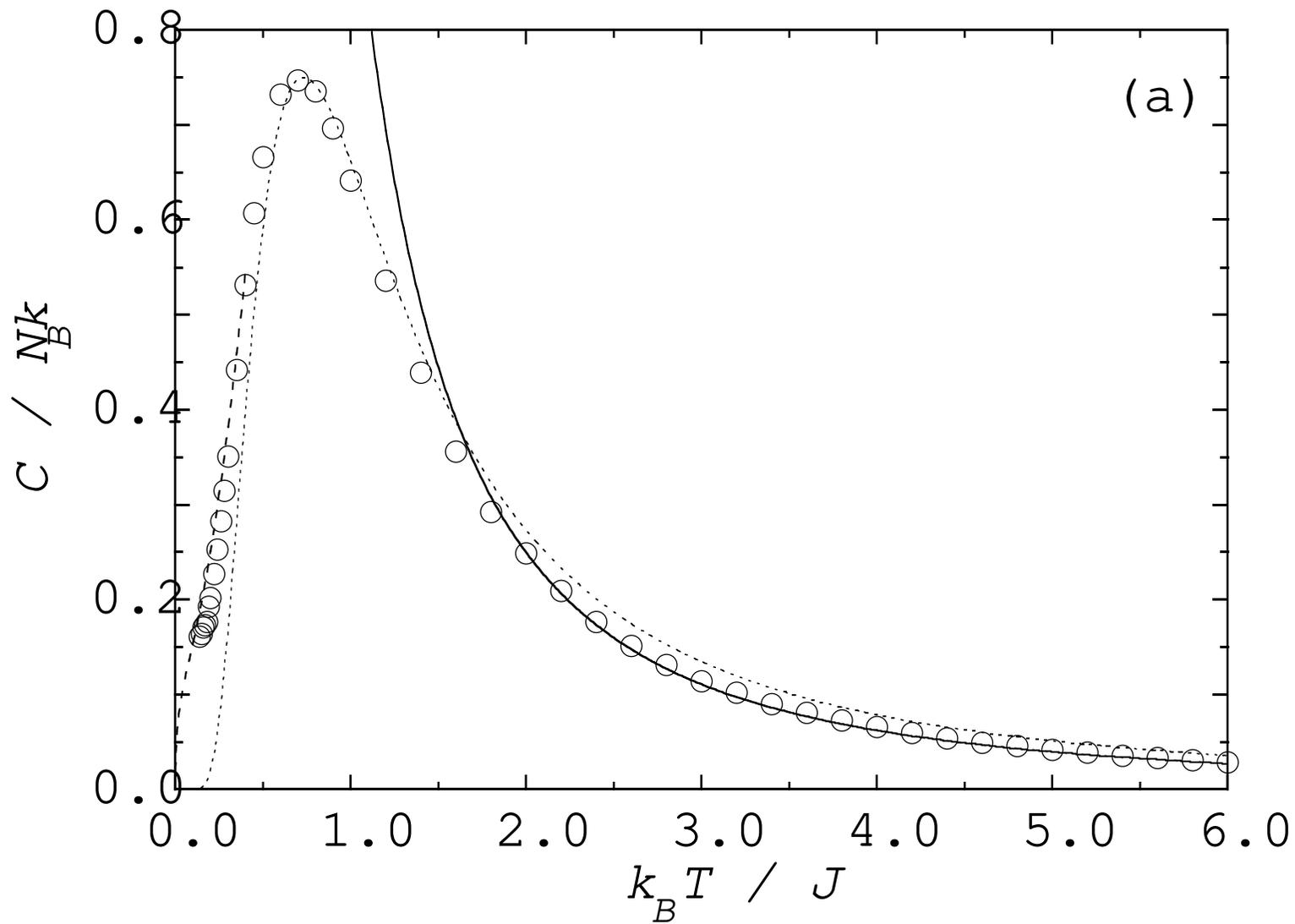
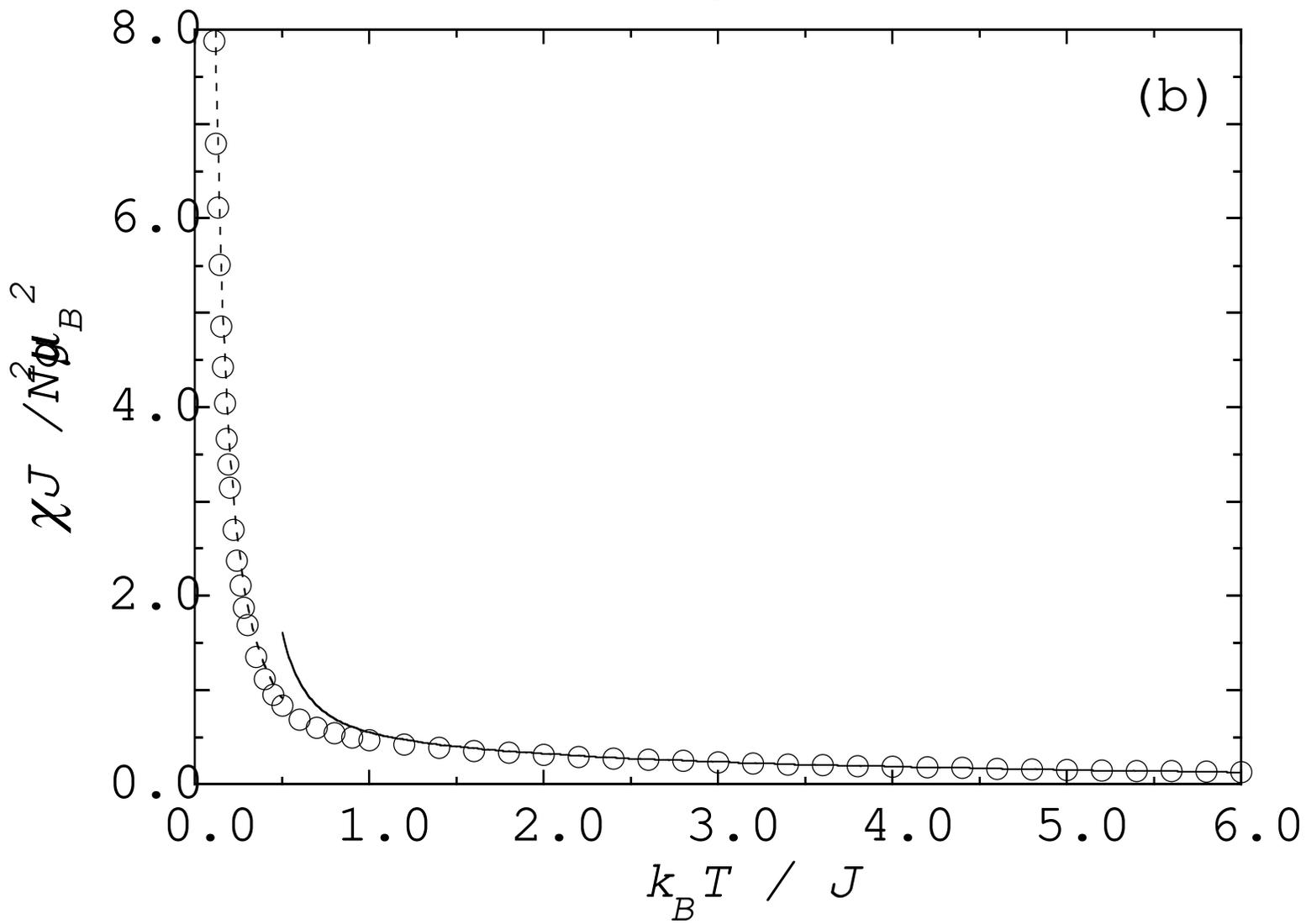
\begin{figure}
%\epsfxsize=70mm %%% 70 is suitable for 2 column
%\centerline{\epsfbox{fig1a.ps}}
%\vspace{5mm}
%\epsfxsize=70mm %%% 70 is suitable for 2 column 
%\centerline{\epsfbox{fig1b.ps}} 
\caption{Quantum Monte Carlo Calculation ($\bigcirc$) of the specific
         heat (a) and the magnetic susceptibility (b) as a function of
         temperature for the Heisenberg ferrimagnetic spin chain with
         alternating spins $1$ and $1/2$ of $N=32$.
         The solid lines: High-temperature series-expansion results for
         the present model within the up-to-$(J/k_{\rm B}T)^3$
         approximation.
         The dashed lines: Low-temperature Behavior [15] of the
         spin-$1/2$ ferromagnetic Heisenberg chain obtained by
         solving the thermodynamic Bethe-ansatz integral equations,
         where the thermal quantities are expanded by powers of
         $(k_{\rm B}T/J)^{1/2}$.
         The specific heat and the magnetic susceptibility are plotted
         within the up-to-$(k_{\rm B}T/J)^{3/2}$ and the
         up-to-$(J/k_{\rm B}T)$ approximations, respectively.
         The dotted line: The Schottky-type specific heat with its gap
         being set equal to $1.759J$, that is, the energy difference
         between the ground state and the antiferromagnetic excitation
         branch of the present model.}
\label{F:QMC}
\end{figure}

   We show in Fig. \ref{F:QMC} temperature dependences of the specific
heat $C$ (a) and the magnetic susceptibility $\chi$ (b).
Although the overall temperature dependences of the recent DMRG
findings \cite{Pati1} are similar to ones of our QMC results, the two
calculations are not in quantitative agreement with each other.
We have confirmed that quantum transfer-matrix \cite{Bets1}
calculation for short chains precisely reproduce the present QMC
findings except for very low temperatures, where overshort chains may
pretend to be gapped.
Furthermore high-temperature series-expansion calculation helps us
to verify our numerical treatment.
Within the up-to-$t^{-3}$ approximation, the specific heat and the
magnetic susceptibility are expanded as
%-------------------------------------------------------------------------
%   high-temperature expansion
%-------------------------------------------------------------------------
\begin{eqnarray}
  \frac{C}{Nk_{\rm B}}
     &=&t^{{\mbox -}2}+O(t^{{\mbox -}4})\,,
  \label{E:HTSE-C} \\%----------------------------------------------------
   \frac{\chi J}{Ng^2\mu_{\rm B}^2}
      &=&\frac{11}{12}t^{{\mbox -}1}
        -\frac{2}{3}t^{{\mbox -}2}
        +\frac{11}{36}t^{{\mbox -}3}
        +O(t^{{\mbox -}4})\,,
   \label{E:HTSE-S} %-----------------------------------------------------
\end{eqnarray}
where $t\equiv k_{\rm B}T/J$ with the Boltzmann constant $k_{\rm B}$.
The asymptotic curves (\ref{E:HTSE-C}) and (\ref{E:HTSE-S}) are also
shown in Fig. \ref{F:QMC}, which convincingly fit the numerical
results.

   The usual spin-wave treatment diagonalizes the
Hamiltonian (\ref{E:H}) with no field as \cite{Pati1,Breh1}
%-------------------------------------------------------------------------
%   spin-wave hamiltonian 
%-------------------------------------------------------------------------
${\cal H}=E_{\rm g}+\sum_k(
 \omega_k^-\alpha_k^\dagger\alpha_k
+\omega_k^+\beta_k^\dagger\beta_k)$,
where
%-------------------------------------------------------------------------
%   spin-wave dispersion
%-------------------------------------------------------------------------
$\omega_k^{\mp}=\omega_k\mp J(S-s)$ with
$\omega_k=J[(S-s)^2+4Ss\sin^2(ak)]^{\frac{1}{2}}$,
$E_{\rm g}\equiv E_0+E_1$ is the ground state energy with
$E_0=-2JSsNJ$ and $E_1=\sum_k[\omega_k-J(S+s)]$, 
and $a$ is the lattice spacing. 
$\alpha_k^\dagger$ and $\beta_k^\dagger$ are the creation
operators of the ferromagnetic and the antiferromagnetic spin waves
with momentum $k$.
The spin-$S$ ferromagnetic Heisenberg chain
exhibits the spin-wave excitations with a quadratic dispersion
$\omega_k=2JS\left[1-\cos(ak)\right].$
Thus, only in the $S=2s$ cases, the ferromagnetic branch of the
spin-$(S,s)$ ferrimagnets show exactly the same dispersion as the
spin-$s$ ferromagnets exhibit at small momenta in the unit of the
unit-cell length being unity.
Hence we expect the present model to behave like the spin-$1/2$
ferromagnet at low temperatures.

   The precise low-temperature behavior of the spin-$1/2$ ferromagnet
has been revealed by Takahashi and Yamada \cite{Taka1}.
Numerically solving the thermodynamic Bethe-ansatz integral equations,
they succeeded in expanding the thermal quantities by powers of
$t^{1/2}$ as
%-------------------------------------------------------------------------
%   Bethe ansatz 
%-------------------------------------------------------------------------
\begin{eqnarray}
   \frac{C}{Nk_{\rm B}}
      &=&0.7815 t^{\frac{1}{2}}
        -2.00 t
        +3.5  t^{\frac{3}{2}}
        +O(t^2)\,,
   \label{E:BA-C} \\%-----------------------------------------------------
   \frac{\chi J}{Ng^2\mu_{\rm B}^2}
      &=&0.04167t^{{\mbox -}2}
        +0.145  t^{{\mbox -}\frac{3}{2}}
        +0.17   t^{{\mbox -}1}
        +O(t^{{\mbox -}\frac{1}{2}})\,,
   \label{E:BA-S}%--------------------------------------------------------
\end{eqnarray}
which are also plotted in Fig. \ref{F:QMC}.
Although the QMC calculation can not reach low enough temperatures, yet
our findings allow us to conclude that {\it the present model is
identified with the spin-$1/2$ ferromagnet at low enough temperatures.}
We note that the lowest-temperature QMC estimates, which
successfully imply the $T^{1/2}$ asymptotic behavior of the specific
heat and the $T^{-2}$ divergence of the magnetic susceptibility, were
obtained through the improved algorithm \cite{Yama3} by spending
forty million MC steps on each data point.

    At mid temperatures in the specific heat,
the antiferromagnetic aspect most clearly appears. 
The specific heat exhibits a sharp peak, rather than a broad one
characteristic of ferromagnets, at $k_{\rm B}T/J\simeq 0.74$ and
therefore reminds us of the Schottky anomaly peculiar to the
antiferromagnetic specific heat \cite{Yama3,Bets1,Blot1,Yama4}.
It is interesting to fit the QMC result to the Schottky-type
specific heat
%-------------------------------------------------------------------------
%   Schottky
%-------------------------------------------------------------------------
\begin{equation}
   \frac{C}{Nk_{\rm B}}
      =A\left(\frac{\Delta}{2k_{\rm B}T}\right)^2
       {\rm sech}^2\left(\frac{\Delta}{2k_{\rm B}T}\right)\,,
\label{E:Schottky}
\end{equation}
with $\Delta$ being set equal to the excitation gap to the
antiferromagnetic branch of the present model, $1.759J$
\cite{Breh1,Yama2}, and a fitting parameter $A$.
We find a fine fit with $A=1.7$ as shown in Fig. \ref{F:QMC} and thus
recognize that {\it the mid-temperature behavior of the specific heat is
well attributed to the gapped antiferromagnetic excitations.}

%-------------------------------------------------------------------------
%                        MODIFIED SPIN-WAVE
%-------------------------------------------------------------------------
   Now the ferromagnetic and the antiferromagnetic aspects of the model
are both revealed.
We inquire further into this picture developing the modified spin-wave
(MSW) theory \cite{Taka2}.
Introducing the additional constraint of the total
magnetization being zero into the theory, 
Takahashi \cite{Taka3} not
only overcame the difficulty in the conventional spin-wave theory
but also succeeded in correctly evaluating
the low-temperature behavior of various thermal quantities.
His idea was further applied to the two-dimensional isotropic
antiferromagnets \cite{Taka4,Hirs2} and thus opened the way for a
quantitative argument of low-dimensional magnets in terms of the
spin-wave picture.
At finite temperatures, we replace $\alpha_k^\dagger\alpha_k$ and
$\beta_k^\dagger\beta_k$ in the spin-wave Hamiltonian by
$\widetilde n^\pm_k\equiv\sum_{n^-,n^+}n^\pm P_k(n^-,n^+)$,
where
$P_k(n^-,n^+)$ is the probability of $n^-$ ferromagnetic and 
$n^+$ antiferromagnetic spin waves appearing in the $k$-momentum 
state and
satisfies $\sum_{n^-,n^+} P_k(n^-,n^+)=1$ for all $k$'s.
Then the free energy at zero field is expressed as
%-------------------------------------------------------------------------
%   free energy 
%-------------------------------------------------------------------------
\begin{eqnarray}
   F&=&E_{\rm g}
      +\sum_k
        (\widetilde n^-_k\omega_k^-+\widetilde n^+_k\omega_k^+)
      \nonumber\\
    &+&k_{\rm B}T
      \sum_k\sum_{n^-,n^+}P_k(n^-,n^+){\rm ln}P_k(n^-,n^+)\,.
   \label{E:F}%-----------------------------------------------------------
\end{eqnarray}
First, we consider minimization of the free energy (\ref{E:F}) with
respect to $P_k(n^-,n^+)$'s under the condition of zero magnetization,
%-------------------------------------------------------------------------
%   zero-magnetization
%-------------------------------------------------------------------------
\begin{equation}
   \langle S^z+s^z\rangle=
   (S-s)N-\sum_k(\widetilde n^-_k-\widetilde n^+_k)=0\,.
   \label{E:constM}
\end{equation}
The free energy and the magnetic susceptibility at thermal equilibrium
are obtained within a set of self-consistent equations:
%-------------------------------------------------------------------------
%   at thermal equilibrium
%-------------------------------------------------------------------------
\begin{eqnarray}
   &&F=E_{\rm g}
      +\mu(S-s)N
      -k_{\rm B}T\sum_{k}\sum_{\sigma=\pm}
       {\rm ln}(1+\widetilde n^\sigma_k)
     \,,\label{E:MSW1SC-F}\\%---------------------------------------------
   &&\chi=\frac{(g\mu_{\rm B})^2}{3k_{\rm B}T}
          \sum_k\sum_{\sigma=\pm}
          \widetilde n^\sigma_k(1+\widetilde n^\sigma_k)
     \,,\label{E:MSW1SC-S}%-----------------------------------------------
\end{eqnarray}
with
$\widetilde n^\pm_k=
[{\rm e}^{(\omega_k^\pm\pm\mu)/k_{\rm B}T}-1]^{-1}$,
where
$\mu$ is a Lagrange multiplier determined by the condition
(\ref{E:constM}).
The susceptibility has been obtained by calculating the thermal average
of $M^2$ \cite{Taka2}.
Equations (\ref{E:MSW1SC-F}) and (\ref{E:MSW1SC-S}) are expanded in
powers of $t^{1/2}$ at low temperatures and result in
%-------------------------------------------------------------------------
%   low-temperature expansion 
%-------------------------------------------------------------------------
\widetext
\begin{eqnarray}
  && \frac{C}{Nk_{\rm B}}
      =\frac{3}{4}\left(\frac{S-s}{Ss}\right)^{\frac{1}{2}}
         \frac{\zeta(\frac{3}{2})}{\sqrt{2\pi}} t^{\frac{1}{2}}
        -\frac{1}{Ss} t
       +\frac{15}{32(S-s)^{\frac{1}{2}}(Ss)^{\frac{3}{2}}}
         \left[
          \frac{(S^2+Ss+s^2)\zeta(\frac{5}{2})}
              {\sqrt{2\pi}}
         -\frac{4\zeta(\frac{1}{2})}{\sqrt{2\pi}}
         \right]
        t^{\frac{3}{2}}
        +O(t^2)
         \,,\label{E:MSWLTSE-C}\\%----------------------------------------
  && \frac{\chi J}{N(g\mu_{\rm B})^2}
      =\frac{Ss(S-s)^2}{3} t^{{\mbox -}2}
      -(Ss)^{\frac{1}{2}}(S-s)^{\frac{3}{2}}
       \frac{\zeta(\frac{1}{2})}{\sqrt{2\pi}}t^{{\mbox -}\frac{3}{2}}
      +(S-s)\left[\frac{\zeta(\frac{1}{2})}{\sqrt{2\pi}}\right]^2
       t^{{\mbox -}1}
      +O(t^{{\mbox -}\frac{1}{2}})
         \,,\label{E:MSWLTSE-S}%------------------------------------------
\end{eqnarray}
\narrowtext
where $\zeta(z)$ is Riemann's zeta function.
In the case of $(S,s)=(1,1/2)$, the expressions (\ref{E:MSWLTSE-C})
and (\ref{E:MSWLTSE-S}) coincide with Eqs. (\ref{E:BA-C}) and
(\ref{E:BA-S}) up to the order $t$ and the order $t^{-1}$,
respectively, and therefore again show us 
{\it the identity
between the present model and the spin-$1/2$ ferromagnet at low
temperatures.}

   The above-demonstrated MSW approach gives a satisfactory description
at low temperatures, whereas we immediately find that it never applies
away from the low-temperature region.
For ferromagnets, the zero-magnetization constraint is not only
convincing in that the thermal average of the magnetization should
be zero at zero field, but also plays a role of keeping the number of
bosons constant.
This is not the case for ferrimagnets as well as for antiferromagnets
\cite{Taka4,Hirs2}.
Here  the spin-wave treatment with the condition (\ref{E:constM})
still results in the divergence of the number of bosons
at high temperatures.
Can we control the number of bosons keeping the above-obtained
low-temperature behavior unchanged and give a convincing description
in a wider temperature region?
We may answer yes replacing the condition (\ref{E:constM}) by
%-------------------------------------------------------------------------
%   staggered magnetization
%-------------------------------------------------------------------------
\begin{equation}
   \langle :S^z-s^z: \rangle
      =(S+s)N-J(S+s)\sum_k\sum_{\sigma=\pm}
       \frac{\widetilde n^\sigma_k}{\omega_k}
      =0\,,
   \label{E:constSM}
\end{equation}
where the normal ordering is taken with respect to $\alpha$ and
$\beta$.
The spin-wave theory shows us that the classical staggered
magnetization $(S+s)N$ is modified into $(S+s)N-2\tau$ with a quantum
spin reduction $\tau$\cite{Breh1}.
Equation (\ref{E:constSM}) claims that the thermal fluctuation 
of the staggered magnetization be constrained to take the classical value.
This is analogous to Eq.(\ref{E:constM}), which claims that the thermal
fluctuation of the magnetization be the classical magnetization 
$(S-s)N$.
We stress that the constraint (\ref{E:constSM}) leads in fact to
exactly the same expressions as Eqs. (\ref{E:MSWLTSE-C})
and (\ref{E:MSWLTSE-S}) at low temperatures.
Now we again obtain a set of self-consistent equations:
%-------------------------------------------------------------------------
%   at thermal equilibrium
%-------------------------------------------------------------------------
\begin{eqnarray}
   &&F=E_{\rm g}
      +\mu(S+s)N
      -k_{\rm B}T\sum_{k}\sum_{\sigma=\pm}
       {\rm ln}(1+\widetilde n^\sigma_k)
     \,,\label{E:MSW2SC-F}\\%---------------------------------------------
   &&\chi=\frac{(g\mu_{\rm B})^2}{3k_{\rm B}T}
          \sum_k\sum_{\sigma=\pm}
          \widetilde n^\sigma_k(1+\widetilde n^\sigma_k)
     \,,\label{E:MSW2SC-S}%-----------------------------------------------
\end{eqnarray}
with 
$\widetilde n^\pm_k=[{\rm e}^{\left(
\omega_k^\pm\omega_k-\mu J(S+s)\right)/\omega_kk_{\rm B}T}-1]^{-1}$,
where $\mu$ is a Lagrange multiplier due to the condition
(\ref{E:constSM}).

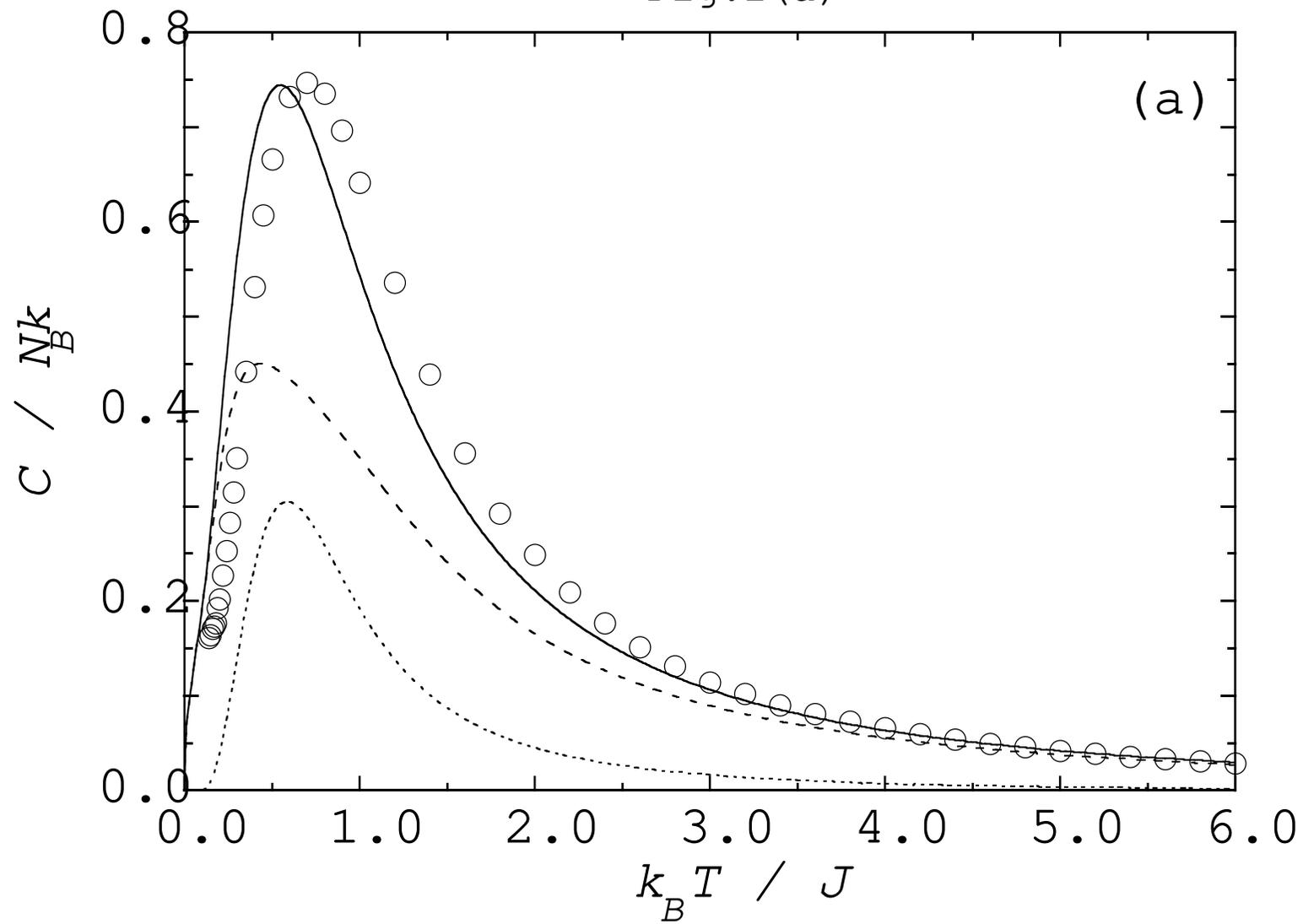
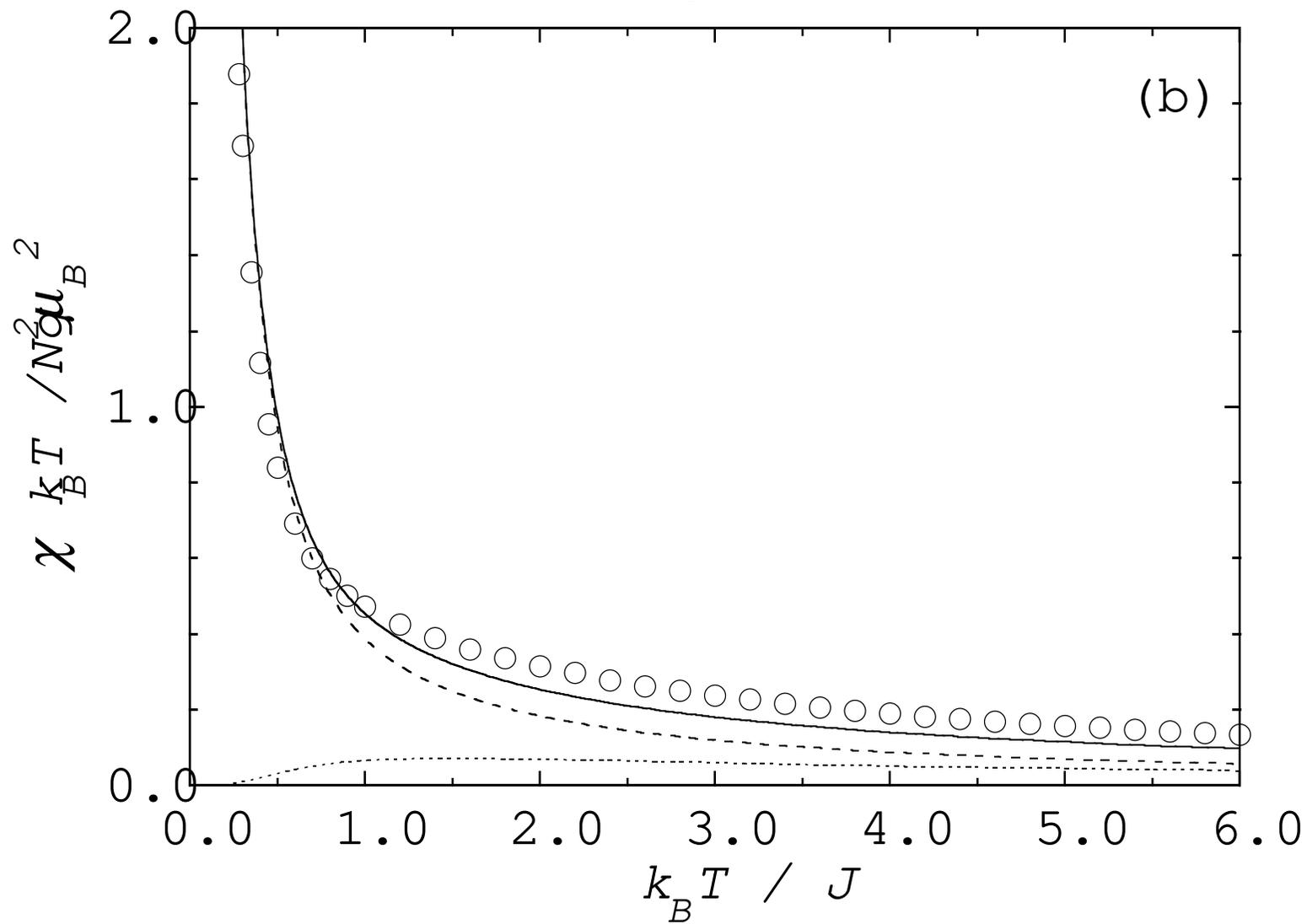
\begin{figure}
%\epsfxsize=70mm %%% 70 is suitable for 2 column
%\centerline{\epsfbox{fig2a.ps}}
%\vspace{5mm}
%\epsfxsize=70mm %%% 70 is suitable for 2 column 
%\centerline{\epsfbox{fig2b.ps}} 
\caption{Modified spin-wave calculation (solid lines) of the specific
         heat (a) and the magnetic susceptibility (b) as a function of
         temperature for the Heisenberg ferrimagnetic spin chain with
         alternating spins $1$ and $1/2$ in the thermodynamic limit,
         where both ferromagnetic and antiferromagnetic spin waves
         contribute to the thermal quantities.
         We effectively extract each contribution of the ferromagnetic
         and the antiferromagnetic spin waves and show them by dashed
         and dotted lines, respectively.
         Quantum Monte Carlo calculation ($\bigcirc$) is again plotted
         for the sake of comparison.}
\label{F:MSW}
\end{figure}

   We numerically solve Eqs. (\ref{E:constSM}) and (\ref{E:MSW2SC-F})
in the thermodynamic limit, 
and visualize them for $(S,s)=(1,1/2)$ in Fig. \ref{F:MSW}, where the QMC
estimates are shown again.
We find that the MSW calculation not only correctly describes the actual
behaviors at low enough temperatures but also well reproduces the
overall temperature dependences.
The $T^{1/2}$ standing up, the Schottky-like peak, the $T^{-2}$ decay
of the specific heat, and the $T^{-2}$ divergence, the $T^{-1}$ decay
of the susceptibility, they are all successfully described by our MSW
approach.
However, the spin-wave excitations underestimate the peak temperature
of the specific heat.
This is because the spin-wave theory results in the gap
$\omega_{k=0}^+=J$, which is smaller than the true value $1.759J$.
In order to separately observe the contributions of the ferromagnetic
and the antiferromagnetic modes, we regard
%-------------------------------------------------------------------------
%   each contribution
%-------------------------------------------------------------------------
$F_{\rm AF}\equiv -k_{\rm B}T\sum_k{\rm ln}(1+\widetilde n_k^+)$ and
$\chi_{\rm AF}\equiv (g\mu_{\rm B})^2(3k_{\rm B}T)^{-1}
 \sum_k\widetilde n_k^+(1+\widetilde n_k^+)$
as the antiferromagnetic contribution, while we define the
ferromagnetic background as
$F_{\rm F}\equiv F-F_{\rm AF}$ and
$\chi_{\rm F}\equiv \chi-\chi_{\rm AF}$.
Each contribution in the specific heat is numerically calculated from
$F_{\rm F}$ and $F_{\rm AF}$, respectively.
We show in Fig. \ref{F:MSW} the thus-obtained separate contributions.
We clearly observe that $C_{\rm F}\rightarrow C$ and
$\chi_{\rm F}\rightarrow\chi$ as $T\rightarrow 0$.
On the other hand, due to the excitation gap, $C_{\rm AF}$ and
$\chi_{\rm AF}$ exponentially vanish as $T\rightarrow 0$.
{\it It is due to the antiferromagnetic mode that the Schottky-like
peak appears}.
Finally, while the above consideration is enlightening, we admit
that the present definition for $C_{\rm F}$ and $C_{\rm AF}$ may not
be relevant at high temperatures, where the $\mu$-term in Eq.
(\ref{E:MSW2SC-F}) should not simply be incorporated into the
ferromagnetic part.
Such a ambiguity inevitably occurs because here the thermal quantities
are obtained through the nonlinear equations.

   We have investigated
thermodynamic properties of the ferrimagnetic mixed-spin chains and
revealed that the ferromagnetic and the antiferromagnetic aspects
simultaneously lie in the model.
It may also be emphasized that the spin-wave theory has so successfully
been applied to the model of one dimension as to describe its whole
thermal behavior.
Not only direct observation of the thermal quantities but also
neutron-scattering measurements of the antiferromagnetic excitations
are fascinating experiments worth trying.

%%\acknowledgments

%-------------------------------------------------------------------------
%   acknowledgments
%-------------------------------------------------------------------------
   The authors would like to thank H.-J. Mikeska, S. Brehmer, and
S. K. Pati for their useful comments and fruitful discussions.
This work was supported by the Japanese Ministry of Education, Science,
and Culture through the Grant-in-Aid 09740286 and by a Grant-in-Aid
from the Okayama Foundation for Science and Technology.
Most of the numerical computation was done using the facility of the
Supercomputer Center, Institute for Solid State Physics, University of
Tokyo.

%-------------------------------------------------------------------------
%   figure 1
%-------------------------------------------------------------------------

%-------------------------------------------------------------------------
%   figure 2
%-------------------------------------------------------------------------

\widetext

\begin{references}

\bibitem{Hald1}
   F. D. M. Haldane,
      Phys. Lett. {\bf 93A}, 464 (1983);
      Phys. Rev. Lett. {\bf 50}, 1153 (1983).

\bibitem{Fuku1}
   T. Fukui and N. Kawakami,
      Phys. Rev. B {\bf 55}, 14709 (1997);
      {\it ibid.} {\bf 56}, 8799 (1997).

\bibitem{Alca1}
   F. C. Alcaraz and A. L. Malvezzi,
      J. Phys.: Math. Gen. {\bf 30}, 767 (1997).

\bibitem{Pati1}
   S. K. Pati, S. Ramasesha, and D. Sen,
      Phys. Rev. B {\bf 55}, 8894 (1997);
      preprint (cond-mat/9704057).

\bibitem{Breh1}
   S. Brehmer, H.-J. Mikeska, and S. Yamamoto,
      J. Phys.: Condens. Matter {\bf 9}, 3921 (1997).

\bibitem{Yama1}
   S. Yamamoto,
      Int. J. Mod. Phys. C {\bf 8}, 609 (1997).

\bibitem{Nigg1}
   H. Niggemann, G. Uimin, and J. Zittartz,
      J. Phys.: Condens. Matter {\bf 9}, 9031 (1997).

\bibitem{Yama2}
   S. Yamamoto, H.-J. Mikeska, and S. Brehmer,
      unpublished.

\bibitem{Kahn1}
   O. Kahn, Y. Pei, and Y. Journaux,
      in {\it Inorganic Materials}, edited by D. W. Bruce and D. O'Hare
      (John Wiley \& Sons, New York, 1992), p. 95.

\bibitem{Lieb1}
  E. Lieb and D. Mattis,
      J. Math. Phys. {\bf 3}, 749 (1962).

\bibitem{Suzu1}
   M. Suzuki,
      Prog. Theor. Phys. {\bf 56}, 1454 (1976).

\bibitem{Hirs1}
   J. E. Hirsch, R. L. Sugar, D. J. Scalapino, R. Blankenbecler,
      Phys. Rev. B {\bf 26}, 5033 (1982).

\bibitem{Yama3}
   S. Yamamoto,
      J. Phys. Soc. Jpn. {\bf 64}, 4051 (1995);
      Phys. Rev. B {\bf 53}, 3364 (1996).

\bibitem{Bets1}
   H. Betsuyaku,
      Prog. Theor. Phys. {\bf 73}, 319 (1985);
   H. Betsuyaku and T. Yokota,
      {\it ibid.} {\bf 75}, 808 (1986).

\bibitem{Taka1}
   M. Takahashi and M. Yamada,
      J. Phys. Soc. Jpn. {\bf 54}, 2808 (1985);
   M. Yamada and M. Takahashi,
      {\it ibid.} {\bf 55}, 2024 (1986).

\bibitem{Blot1}
   H. W. J. Bo\"ote,
      Physica {\bf 78}, 302 (1974);
      Physica {\bf 79B}, 427 (1975).

\bibitem{Yama4}
   S. Yamamoto and S. Miyashita,
      Phys. Rev. B {\bf 48}, 9528 (1993).

\bibitem{Taka2}
   M. Takahashi,
      Prog. Theor. Phys. Suppl. {\bf 87}, 233 (1986).

\bibitem{Taka3}
   M. Takahashi, Phys. Rev. Lett. {\bf 58}, 168 (1987).

\bibitem{Taka4}
   M. Takahashi, Phys. Rev. B {\bf 40}, 2494 (1989).

\bibitem{Hirs2}
   J. E. Hirsch and S. Tang,
      Phys. Rev. B {\bf 40}, 4769 (1989);
   S. Tang, M. E. Lazzouni, and J. E. Hirsch,
      Phys. Rev. B {\bf 40}, 5000 (1989).

\end{references}
\end{document}